\documentclass[pre-print,showpacs,amsmath,amssymb]{revtex4}

\usepackage{graphicx}
\usepackage{dcolumn}
\usepackage{bm}

\begin{document}

\title{A two measure model of dark energy and dark matter}
\author{Eduardo Guendelman}
\affiliation{Department of Physics, Ben-Gurion University, Beer-Sheva, Israel\\
Physics Department, CSU Fresno, Fresno, CA 93740 USA}
 \email{guendel@bgu.ac.il}

\author{Douglas Singleton}
\affiliation{Physics Department, CSU Fresno, Fresno, CA 93740 USA \\
and \\
Institut f{\"u}r Mathematik, Universit{\"a}t Potsdam
Am Neuen Palais 10, D-14469 Potsdam, Germany}
\email{dougs@csufresno.edu}

\author{N. Yongram}
 \altaffiliation{Supported by Thailand Center of Excellence in Physics (ThEP)}
\affiliation{ThEP's CRL, NEP, The Institute of Fundamental Study (IF)\\
Naresuan\;University,\;Phitsanulok\;65000,\;Thailand\\
Thailand Center of Excellence in Physics\\
CHE, Ministry of Education, Bangkok 10400, Thailand\\
and \\
Physics Department, CSU Fresno, Fresno, CA 93740 USA
\footnote{As Visiting Researcher}}
\email{nattapongy@nu.ac.th}
 
\date{\today}

\begin{abstract}
In this work we construct a unified model of dark energy and dark matter. This is done
with the following three elements: a gravitating scalar field, $\phi$ with a non-conventional 
kinetic term, as in the string theory tachyon; an arbitrary potential, $V(\phi)$;
two measures -- a metric measure ($\sqrt{-g}$) and a non-metric measure
($\Phi$). The model has two interesting features: 
(i) For potentials which are unstable and would give rise to tachyonic scalar field, this model
can stabilize the scalar field. (ii) The form of the dark energy and dark matter that results from this model
is fairly insensitive to the exact form of the scalar field potential.        
\end{abstract}


\maketitle

\section{\label{Sec01}Introduction}

Currently, observations point to the Universe being dominated by two very different types of non-luminous or ``dark"
fluids -- dark energy, which leads to gravitational repulsion, and dark matter, which is gravitational attractive. 
Dark energy was observationally discovered rather recently through the observation of type Ia supernova \cite{riess}
\cite{perlmutter}. Dark matter was first postulated in the 1930s, separately by J. Oort and F. Zwicky, due to the anomaly of the
orbital velocity of some stars in the Milky Way galaxy and the orbital velocity of galaxies in clusters. 
A recent review of dark matter is given in \cite{garrett}, reviews of dark energy can be found in \cite{starobinsky} 
and a review of both dark matter and dark energy can be found in \cite{sahni}.  

In this paper we study a simple model which has two different types of measures -- a metric and non-metric
measure. The introduction of the non-metric measure opens the possibility, in some cases, to reverse the expected 
stability of the scalar field. For a scalar potential, $V(\phi )$, which is naively stable, our model can lead to 
an unstable scalar field, while for a potential, $V(\phi )$, which is naively unstable our model
can lead to a stable scalar field. Thus our model might be able to stabilize systems which are unstable
due to the presence of tachyons. String theory contains tachyons \cite{Sen} which have been studied in the 
context of cosmology \cite{Feinstein} \cite{gibbons} \cite{frolov}. Generally, the string theory tachyon can 
be problematic for cosmological models since it can lead to instabilities. 

Another feature of our model is that the scalar field can act both as dark energy and dark matter.   
This dual behavior of the scalar field is largely independent of form of the potential, $V(\phi )$. This combined 
treatment of dark energy and dark matter is similar to Chaplygin gas models \cite{kamenshchik}.
Reference \cite{bilic} gave a unification of dark matter and dark energy via a Chaplygin gas which is
close to the unified treatment that we present below in terms of two measure theory, but in our case 
the effect of the modified measure produces two important new effects: 1) the appearance of an integration 
constant that  makes the observed vacuum energy density totally decoupled from the parameters of the Lagrangian 
and 2) the resulting DE-DM unified theory resembles much more the traditional Lambda - cold dark matter model, 
not only for the homogeneous solutions, but also for the perturbations and therefore the corresponding structure 
formation picture. Finally, since in this model the dark energy comes from a dynamical scalar field this leads to 
dynamical dark energy \cite{alam} as opposed to the simple case where dark energy is associated with a non-dynamical 
cosmological constant.

The starting point is to introduce a new measure, $\Phi$, which is independent of 
the metric, $g_{\mu\nu}$. This additional measure can be constructed from four scalar fields 
($\varphi ^i$ where $i=1,2,3,4$) for example
\begin{equation}\label{eq01}
\Phi=\varepsilon^{\mu\nu\alpha\beta}\varepsilon_{ijkl}\partial_{\mu}\varphi^{i}
\partial_{\nu}\varphi^{j}\partial_{\alpha}\varphi^{k}\partial_{\beta}\varphi^{l}.
\end{equation} 
In such two measures theories \cite{Guendelman_Kaganovich}, the general action can be written as
\begin{equation}
S=\int{\cal L}_{1}\sqrt{-g}\;\mathrm{d}^{4}x +\int {\cal L}_{2}\Phi\mathrm{d}^{4}x\label{eq02}
\end{equation}
where ${\cal L}_{1}$ and ${\cal L}_{2}$ are $\varphi^{i}$-independent. A similar type of
non-metric structure is used in \cite{rajpoot} to study supersymmetric models. 

The introduction of $\Phi$ in conjunction with $\sqrt{-g} = \sqrt{-\det(g_{\mu\nu})}$ as measures of integration provides many 
uses -- new approaches to scale invariance and its breaking \cite{Guendelman1999}, 
brane-world scenarios \cite{Guendelman2004}, and non-singular cosmologies \cite{Campo2011}. 

A general feature of these two measure theories is that the equation of motion of the $\varphi^{i}$ fields,
given by 
\begin{equation}
A^{\mu}_{i}\partial_{\mu}{\cal L}_{2}=0 ~, \label{eq03}
\end{equation}
where $A^{\mu}_{i}=\varepsilon^{\mu\nu\alpha\beta}\varepsilon_{ijkl}\partial_{\nu}\varphi^{j}
\partial_{\alpha}\varphi^{k}\partial_{\beta}\varphi^{l}$, leads to 
\begin{equation}
{\cal L}_{2}= \mathrm{constant} \label{eq04}
\end{equation}
It is also interesting to note that the two measures theory action \eqref{eq02} is invariant (up to the integral of a total derivative) 
under the infinite dimensional group of transformations \cite{Guendelman_Kaganovich} ($f^{i}$ being arbitrary functions), 
\begin{equation} \label{symm}
\varphi^{i} \rightarrow \varphi^{i} + f^{i} ({\cal L}_{2}) 
\end{equation}
as long as ${\cal L}_{1}$ and ${\cal L}_{2}$ are $\varphi^{i}$ independent. One can think of this symmetry as the reason the 
action of the two measures theories may be preserved say under quantum corrections.
The two measure theories have many points of similarity with ``Lagrange Multiplier Gravity (LMG)'' \cite{Lim2010,Capozziello2010}. 
In LMG there is a Lagrange multiplier field which enforces the condition that a certain function is zero. In the 
two measure theory this is equivalent to the constraint given in \eqref{eq04}. The two measure model presented here, 
as opposed to the LMG models of \cite{Lim2010,Capozziello2010} provide us with an arbitrary constant of integration, 
which in the model to be discussed here will be identified as the effective vacuum energy. This ``floating vacuum energy" 
is totally decoupled from the original parameters of the action. The two measures theory could also provide a 
theoretical foundation for the more phenomenological approach of the LMG model, since for example the symmetries like  
\eqref{symm} of the two measure theory can justify the analogous structure in the LMG model, 
if one thinks of the two measures theory as the more basic underlying theory.

As a final comment we note that the introduction of constraints like that in \eqref{eq04} 
can cause Dirac fields 
to contribute to dark energy \cite{Guendelman_Kaganovich2006IJMP} or scalar fields to behave like dust \cite{Lim2010}.

\section{\label{Sec02} Gravity plus scalar field and non-metric measure}

The system we consider is gravity with a scalar field having a non-standard kinetic term
and a non-metric measure like the second term in \eqref{eq02}. The explicit action for the model is  
\begin{equation}
S = S_g + S_m = \int\frac{R}{16 \pi G} \sqrt{-g}\;\mathrm{d}^{4}x + 
\int \frac{K(\phi, \partial _\mu \phi)}{\sqrt{-g} } 
\Phi \mathrm{d}^{4}x  +  \int K( \phi, \partial _\mu \phi )\;\mathrm{d}^{4}x ~, \label{eq05}
\end{equation}
where in the second term, $\Phi$, is from \eqref{eq01} and the DBI-like kinetic term is given by  
\begin{equation} \label{eq06}
K (\phi, \partial _\mu \phi )= V(\phi ) \sqrt{-\det(g_{\mu\nu}+\partial_{\mu}\phi\partial_{\nu}\phi)} =
V(\phi ) \sqrt{-g} \sqrt{1 + g^{\mu \nu} \partial_{\mu}\phi\partial_{\nu}\phi}.
\end{equation}
The DBI-like scalar field, $\phi$, should not be confused with the auxiliary fields, $\varphi ^i$, of \eqref{eq01}
which were used to show how one might define the second, non-metric measure, $\Phi$.
Note that in \eqref{eq05} we have split the action differently than in \eqref{eq02}, where the split was between
the metric measure and non-metric measure. In \eqref{eq05} the split is between the gravitational part and matter part. 
The first term in \eqref{eq05} above is the standard 4D gravitational action and is denoted by $S_g$.
The second plus third term are grouped together in $S_m$ as the matter part of the total action.
The second term is the non-metric measure term -- it is the explicit realization of the 
second term in \eqref{eq02}. The last term is the non-standard kinetic scalar field with a scalar
self-interaction potential $V(\phi )$. Actions having the form of this third term from \eqref{eq05} were
used in \cite{Sen} \cite{garousi} to study tachyons in string theory. Also, actions similar to the 
third term in \eqref{eq05} were investigated in k-essence models \cite{tupper} which sought to 
present a unified picture of dark matter and dark energy analogous to what is suggested here in 
terms of two measure theory. Note that the second term in \eqref{eq05} 
is $V(\phi)$ times the ratio of the square root of the determinants of the two natural ``metrics" -- $g_{\mu\nu}$ 
and $g_{\mu\nu}+\partial_{\mu}\phi\partial_{\nu}\phi$ -- that one can define for this system. One can see that
the metric $g_{\mu\nu}+\partial_{\mu}\phi\partial_{\nu}\phi$ can be related to 5D gravity. Defining $x^5 = \phi (x)$
we find that $d x^5 = \partial _\mu \phi dx ^\mu$ which then gives 
\begin{equation}\label{eq07}
ds^2 = g_{\mu \nu} dx^\mu dx ^\nu + dx^5 dx^5 
\rightarrow  g_{\mu \nu} dx^\mu dx ^\nu + (\partial _\mu \phi dx ^\mu ) (\partial _\nu \phi dx ^\nu )~,
\end{equation}
which finally gives the metric $g_{\mu\nu}+\partial_{\mu}\phi\partial_{\nu}\phi$.

In the second term of the action \eqref{eq05} we have coupled the metric independent measure, $\Phi$, to the ratio 
of the determinants of the two natural metrics. This kind of coupling between $\Phi$ and the two ``metrics" was
used in \cite{Guendelman2004} to study brane world models. The DBI-like scalar field $\phi$ is taken to be 
only a function of the cosmic time $t$. We will show that 
there is a constraint on the action \eqref{eq05} that is a specific realization of the general constraint given in 
\eqref{eq04}. This constraint has the effect of stabilizing unstable scalar fields like the tachyon found in
string theory. In addition we will find that the action in \eqref{eq05} leads to both dark energy and dark matter 
effects, thus giving a unified model of dark energy and dark matter. There are other unified models, such as 
\cite{chaves}, which uses graded Lie algebras to give a unified model of dark energy and dark matter. In the
present proposal the unification of these two ``dark" cosmological elements comes from using two measures -- one
geometric and one non-geometric.      
 
The action in \eqref{eq05} can be seen to fit the two measure structure in \eqref{eq02} if we take 
${\cal L}_{1}=\frac{R}{16 \pi G} + \frac{K}{\sqrt{-g}}$ and ${\cal L}_{2}= \frac{K}{\sqrt{-g}}$. 
Using \eqref{eq03} and \eqref{eq04} this then implies that $\frac{K}{\sqrt{-g}}$ is a constant. This is the
constraint. Using the assumption that $\phi$ is only a function of $t$ and that $g_{0i}=0$ we find that the 
constraint from \eqref{eq04} gives 
\begin{equation}
\label{eq08}
\frac{K}{\sqrt{-g}}= V(\phi ) \sqrt{1+\frac{\dot{\phi}^{2}}{g_{00}}}=M
\end{equation}
where $M$ is a constant. This last equation can be written in the form of an energy equation  
\begin{equation}
\label{eq09}
-\frac{{\dot \phi }^2}{g_{00}} +\frac{M^2}{V^2 (\phi )} = 1 ~.
\end{equation}
This dynamical equation for $\phi$ is of the form of an energy equation $KE + PE = constant$. 
We are using a signature where $g_{00} <0$ so that the first term in \eqref{eq09} is a standard, positive
kinetic term. Because of the positivity of the first term on the left hand side of \eqref{eq09} one
finds that $\frac{M^2}{V^2 (\phi )}$ has the bound
\begin{equation}
\label{eq09a}
\frac{M^2}{V^2 (\phi )} \le  1 \implies |M| \le |V (\phi ) |
\end{equation}
The equality holds when ${\dot \phi } =0$.

From \eqref{eq09} one can define an effective potential $V_{eff} = \frac{M^2}{V ^2 (\phi ) }$ which has the inverse
behavior of the original potential $V (\phi )$. This feature, that the effective potential is the inverse, squaredg   of the
original potential opens up the possibility to change the stability of the system. For example, in reference \cite{Kutasov2003}
the potential $V(\phi) \propto \frac{1}{\cosh(\phi)}$ is studied and is found to lead 
to a tachyonic scalar field and instability. From \eqref{eq09}
this form of the potential leads to an effective potential of the form $V_{eff} (\phi ) \propto [\cosh(\phi)] ^2$ 
which is naively stable and for which one would not expect the scalar field to be tachyonic. 

Now varying the action in \eqref{eq05} with respect to the inverse metric $g^{\mu\nu}$ gives the 
field equations (using either first order or second order formalism)
\begin{equation}\label{eq10}
-\frac{1}{8 \pi G}\left(R_{\mu\nu}-\frac{1}{2}g_{\mu\nu}R\right)+T_{\mu\nu}=0 ~.
\end{equation}
$R_{\mu\nu}$ is the Ricci tensor, $R=g^{\mu\nu}R_{\mu\nu}$ is the Ricci scalar, and $T_{\mu\nu}$ 
is the total energy-momentum tensor. Below we will split the energy-momentum tensor into a term 
associated with the new measure $\Phi$ and a term associated with the scalar field, $\phi$
i.e. $T_{\mu \nu} = T^{\Phi}_{\mu \nu} + T^{\phi}_{\mu \nu}$.

The above considerations will be applied in the framework of Friedmann-Robertson-Walker (FRW)
cosmology with the metric taken to be of the form
\begin{equation} \label{eq11}
ds^{2}=g_{00}dt^{2}+a^{2}(t)\left[\frac{dr^{2}}{1-kr^{2}}+r^{2} (d \theta ^2 + \sin ^2 \theta d \phi ^2) \right] ~.
\end{equation} 
In the above equation $k=-1,0,1$, $g_{\mu\nu}$ has the signature $\mathrm{diag}(-,+,+,+)$, $a(t)$  
is the scale factor as a function of cosmic time $t$, and we have taken the time coordinate $t$
such that the lapse function, $g_{00}$ is not simply $g_{00}=-1$, but may be some other negative constant or may even 
be a function of $t$ i.e. $g_{00} (t)$. When $g_{00}$ is some arbitrary constant different from $-1$ this
just represents a simple re-scaling of the cosmic time. We have kept the explicit $g_{00}$ dependence in order 
to calculate correctly the contribution of the $\Phi$ term to the energy density. 

Now we write down the total energy-momentum tensor from \eqref{eq07} using the definition 
$T_{\mu\nu}=-\frac{2}{\sqrt{-g}}\frac{\delta S_m}{\delta g^{\mu\nu}}$ ($S_m$ is the second
and third terms of the action \eqref{eq05} -- the matter part of the action) 
\begin{equation}
\label{eq12}
T_{\mu \nu} = - \left( 1 + \frac{\Phi}{\sqrt{-g}} \right)\frac{V (\phi )}{\sqrt{1+ g^{\mu \nu} \partial _\mu \phi \partial _\nu \phi }} 
\partial _\mu \phi \partial _\nu \phi + g_{\mu \nu}  V(\phi ) \sqrt{1+ g^{\mu \nu} \partial _\mu \phi \partial _\nu \phi } ~.
\end{equation}
We can break up \eqref{eq12} in terms of the separate energy density and pressures components as
\begin{eqnarray} \label{eq13} 
T_{00} &=& -g_{00} \left( 1 + \frac{\Phi}{\sqrt{-g}} \right) \frac{V(\phi)}{\sqrt{1+ \frac{{\dot \phi }^2}{g_{00}}}} 
\frac{{\dot \phi}^2}{g_{00}}  + g_{00}  V(\phi ) \sqrt{1+ \frac{{\dot \phi }^2}{g_{00}}} = g_{00} \rho  \\
T_{ij} &=&  g_{ij} V(\phi) \sqrt{1+ \frac{{\dot \phi }^2}{g_{00}}}  = - g_{ij} p ~, 
\label{eq13a}
\end{eqnarray}
where we have taken into account that $\phi (t)$ only depends on $t$ so that the partial derivatives of $\phi$ reduce
simply to time derivatives. From \eqref{eq13} \eqref{eq13a} the total energy density and the total pressure are given by 
\begin{eqnarray}
\rho_{\mathrm{total}}&=&\frac{\Phi V^2 (\phi )}{M\sqrt{-g}} \left( 1-\frac{M^2}{V^2 (\phi )} \right) + \frac{V^2 (\phi)}{M}
= \rho ^{\Phi} + \rho ^{\phi} \label{eq14} \\
p_{\mathrm{total}}&=&- M = p^{\phi} ~, \label{eq15}
\end{eqnarray}
where we used \eqref{eq09} in going from \eqref{eq13} \eqref{eq13a} to \eqref{eq14} \eqref{eq15} and we have split the 
energy density and pressure into separate contributions coming from $\Phi$ and $\phi$. The $\Phi$ term does not contribute 
to the pressure i.e. $p ^{\Phi} =0$. The $\phi$ terms contributes both to the energy density and the pressure. 

\subsection{Equations of Motion and solution}

In order to find an explicit expression for $\rho_{\mathrm{total}}$ in \eqref{eq14}  
we need to solve for $\Phi$. This is done using the gauge $g_{00} =-1$ and the 
Euler-Lagrange equations coming from the action 
$S=\int \left( \dfrac{R}{16 \pi G} \sqrt{-g}+{\cal L}_m \right) \mathrm{d}^{4}x$ where
\begin{equation}
\label{eq16}
{\cal L}_m= \frac{K}{\sqrt{-g}} \Phi + V ( \phi ) \sqrt{-g}\sqrt{1- \dot{\phi}^{2}}
= V ( \phi ) \Phi \sqrt{1- \dot{\phi}^{2}}  + V ( \phi ) \sqrt{-g}\sqrt{1- \dot{\phi}^{2}} ~.
\end{equation} 
The Euler-Lagrange equations for this Lagrangian density lead to the equation of motion  
\begin{equation}
\label{eq17}
-\frac{\partial}{\partial{t}}\left(\dfrac{V (\phi) \Phi \dot{\phi}}{\sqrt{1 - \dot{\phi}^{2}}}
+\dfrac{V (\phi ) \sqrt{-g}\dot{\phi}}{\sqrt{1-\dot{\phi}^{2}}}\right)
= \frac{\partial V}{\partial \phi} \left( \Phi \sqrt{1-\dot{\phi}^{2}}
+ \sqrt{-g}\sqrt{1-\dot{\phi}^{2}} \right)  ~.
\end{equation}
It is possible to find a simple closed form expression for $\Phi$ for fairly general $V (\phi )$.
We first note that one can solve \eqref{eq17} for the trivial case when $V (\phi ) = const$. Second
it is also possible to solve \eqref{eq17} for general varying $V (\phi )$. First we
take the field $\phi$ to satisfy $\dot{\phi} \ne 0$ and $\phi$ monotonic, as we discuss later, 
these restrictions can be dropped. 
Using the chain rule and taking into account the assumed conditions on $\phi$ we can write  
\begin{equation}\label{eq18}
\dfrac{\mathrm{d}\;}{\mathrm{d}t} =\dot{\phi}\dfrac{\mathrm{d}\;\;}{\mathrm{d}\phi} ~. 
\end{equation} 
Using \eqref{eq18} allows us to change the derivative on the left hand side of \eqref{eq17} from a 
derivative of $t$ to a derivative of $\phi$. Finally using \eqref{eq08} and \eqref{eq09}
to write $\sqrt{1- \dot{\phi}^{2}} =M/V$ and $\dot{\phi} = \sqrt{1-M^2/V^2}$
allows us to re-write \eqref{eq17} as
\begin{equation}\label{eq19}
\dfrac{\mathrm{d}\;}{\mathrm{d}\phi}\left(\dfrac{\psi V^{2}}{M}\sqrt{1-\dfrac{M^{2}}{V^{2}}}\right) 
= -\left(\dfrac{\mathrm{d}V\;}{\mathrm{d}\phi}\right)\left(\dfrac{\psi M}{V\sqrt{1-\dfrac{M^{2}}{V^{2}}}}\right) ~. 
\end{equation} 
In the above we have defined $\psi=\Phi+\sqrt{-g}$. 

First we consider the simple case when $V (\phi) = const$ so that the right hand side of \eqref{eq19} vanishes.
In this case one immediately sees that $\psi =const$ which then leads to
\begin{equation} \label{eq19a}
\Phi = -\sqrt {-g} + const. ~,
\end{equation}
so that up to a constant the non-metric measure, $\Phi $, is the same as the metric measure, $\sqrt{-g}$.
Without going into the full details, we mention that for this simple case, when one inserts \eqref{eq19a} into the 
equation for the energy momentum tensor \eqref{eq14} \eqref{eq15}, one finds that this leads to a ``dust" contribution 
plus a vacuum energy contribution equal to $M$.

We now treat in detail the less trivial and more interesting case when $V (\phi )$ varies. From \eqref{eq19} we get 
\begin{equation}
\label{eq20}
\int \dfrac{d \psi}{\psi} = -\int\dfrac{2 ~ dV}{V\left(1-\dfrac{M^{2}}{V^{2}}\right)}
\end{equation} 
Carrying out the integration in \eqref{eq20} leads to 
\begin{equation}\label{eq21}
\psi =\dfrac{C (r, \theta ) }{\left( V^2- M^2\right)} ~,
\end{equation} 
where $C (r, \theta )$ is an integration ``constant" which depends on $r$ and $\theta$. 
By substituting the definition $\psi=\Phi+\sqrt{-g}$ in \eqref{eq21}, we obtain
an expression for $\Phi$
\begin{equation}\label{eq22}
\Phi =\dfrac{C (r, \theta ) }{\left(V^2- M^{2}\right)}- \sqrt{-g}~ .
\end{equation}
We will take the time-independent integration function to be
$C (r, \theta) = D r^{2}\sin\theta /\sqrt{1-kr^{2}}$ with $D$ a constant. 
The integration function $C(r,\theta )$ is chosen in this way so as to match and therefore
cancel out the $r$ and $\theta$ dependence of $\sqrt{-g}$ in \eqref{eq14}. This leads to
an energy density which is independent of the spatial coordinates
$r, \theta, \phi$ as required by the homogeneity of FRW space-time.

Finally substituting $\Phi$ from \eqref{eq22} into the expression of the total energy density 
from \eqref{eq14} gives 
\begin{equation}\label{eq23}
\rho_{\mathrm{total}}= M + \dfrac{D}{Ma^{3}}.
\end{equation}
The first term is a constant term (i.e. a cosmological constant-like term) while the second
term has the typical $a^{-3}$ behavior of dust. The pressure for this system is negative and 
is given by \eqref{eq15} as $p_{\mathrm{total}}=- M$.    

We can use this to obtain the equation of state parameter for the system
\begin{equation}
\label{eq24}
w = \frac{p_{total}}{\rho _{total}} = \dfrac{-1}{1+\frac{D}{M^{2} a^3}}  ~,
\end{equation} 
If $D >0$ one can see that $-1<w<0$. At early times (i.e. $a \rightarrow 0$
the denominator in \eqref{eq24} becomes large and $w \rightarrow 0$ from below. At late times
(i.e. $a \rightarrow \infty$) the denominator approaches $-1$ and one effectively has a cosmological constant. 
In this guise our toy model still faces the coincidence problem -- why we happen to live in an 
era where the ordinary matter energy density (i.e. dark matter plus ordinary baryonic matter represented 
by the second term in \eqref{eq23} above) is the same order of magnitude as the ``cosmological constant"
energy density (represented by the first term in \eqref{eq23}). 

Since we have not specified the scalar potential, $V( \phi )$ our results in 
\eqref{eq23} and \eqref{eq24} will occur for a wide range of potentials. Note the constant
energy density term from $\rho_{\mathrm{total}}$ in \eqref{eq23} is not $V (\phi _0)$  (where
$\phi _0$ is the equilibrium value of $\phi $ as determined by \eqref{eq09}) as one would naively expect,
but rather is given by $M$. This ``transformation" of the naive vacuum energy from $V (\phi _0)$ to
$M$ can be traced to the use of the second non-metric measure $\Phi$ and the constraint given by \eqref{eq03}, 
\eqref{eq04} (generically) and \eqref{eq08} (specifically for our model). Thus this model very robustly
leads to dark energy and dust for very general scalar field potentials. Finally, we discuss now, 
and in the next section, how the conditions $\dot{\phi} \ne 0$ and $\phi$ monotonic are not 
really necessary. In fact for an oscillating solution for example, these conditions are satisfied piecewise, 
and for each interval where these conditions hold the derivations above hold. Furthermore the isolated points 
where  $\dot{\phi} = 0$ constitute a set of measure zero and do not contribute to the integration that allowed us 
to solve for the measure $\Phi$. In the next section we will see how the covariant energy momentum 
conservation condition gives the geodesic behavior of inhomogeneous perturbations and also 
gives us another way to look at the equation for the measure $\Phi$, where it will be also evident 
that the isolated points $\dot{\phi} = 0$ do not constitute a problem.

\section{Geodesic behavior for inhomogeneous perturbations}

Before one can claim that the two measure model with scalar field, $\phi$, as described above acts as
dark matter dust, as implied by the second term in \eqref{eq23} one must show that the inhomogeneous perturbations
do in fact propagate like dust i.e. along time-like geodesics. To show this we begin by considering  
the general energy momentum tensor from \eqref{eq12} without the assumptions of homogeneity or isotropy of the space time
which led to the result in \eqref{eq23}. If the perturbations from the homogeneous case studied in the last section
are not too big, we can keep the gradients of the scalar field as time-like vectors -- 
that is we still satisfy $\partial_{\alpha}\phi \partial^{\alpha}\phi < 0$. Then we can express the energy momentum in a 
``fluid form" by defining the four velocity of the fluid as 
\begin{equation}\label{geodesic1}
u_{\mu}=  \frac{\partial_{\mu}\phi}{\sqrt{-\partial_{\alpha}\phi \partial^{\alpha}\phi}}~.
\end{equation}
This four velocity of the fluid is normalized according to
\begin{equation}\label{geodesic2}
u_{\nu}u^{\nu}=  -1
\end{equation}

The energy momentum tensor from \eqref{eq12} now takes the form
\begin{equation}\label{geodesic3}
T_{\mu \nu}= \rho_d u_{\mu}u_{\nu}  + M g_{\mu \nu} ~,
\end{equation}
where we have used the constraint from \eqref{eq08} to write out the $g_{\mu \nu}$ part of 
$T_{\mu \nu}$. The energy density, $\rho _d$, comes from the derivative part of the general
energy momentum tensor in \eqref{eq12} and is defined as 
\begin{equation}
\label{rhod}
\rho_ d =  \left( 1 + \frac{\Phi}{\sqrt{-g}} \right)
\frac{(\partial_{\alpha}\phi \partial^{\alpha}\phi ) V (\phi )}{\sqrt{1+ g^{\mu \nu} \partial _\mu \phi \partial _\nu \phi}}
\end{equation}

We can consider the covariant conservation of the energy momentum tensor from \eqref{geodesic3}. Since the covariant derivatives 
of the metric are zero, the M term is separately conserved. From the covariant conservation of  
the first term, we obtain then
\begin{equation}\label{geodesic4}
\nabla_{\mu}(\rho _d u^{\mu}u^{\nu})= \nabla_{\mu}(\rho _d u^{\mu}) u^{\nu} + \rho _d u^{\mu}\nabla_{\mu} u^{\nu}=0
\end{equation}
We can see now that the two terms in the above equation must vanish separately. This is because the vectors $u^{\nu}$ and 
$ u^{\mu}\nabla_{\mu} u^{\nu}$ are orthogonal and therefore linearly independent.
The orthogonality can be proved by just applying the operator $ u^{\mu}\nabla_{\mu}$ to both sides of the normalization 
condition \eqref{geodesic2}. In fact, according to \eqref{geodesic2}  $u^{\nu}$ is a time-like vector and the  vector 
orthogonal to  it $ u^{\mu}\nabla_{\mu}u^{\nu}$ is space-like, as is most easily seen in the frame 
where  $u^{\nu}= (1,0,0,0)$.
Therefore, in order to satisfy \eqref{geodesic4} one must set to zero both terms separately. 
Therefore we get the two following results that allow us to make a connection to a ``particle" 
interpretation of the model:
\begin{equation}\label{geodesic5}
 \nabla_{\mu}(\rho _d u^{\mu})= 0
\end{equation}
which represents a sort of ``particle number" conservation, and as long as $\rho _d \neq 0 $
(i.e. as long as there are particles at that point in space-time) we obtain,
\begin{equation}\label{geodesic6}
 u^{\mu}\nabla_{\mu} u^{\nu} =0
\end{equation}
which is indeed the geodesic equation for the perturbations or equivalently the geodesic equation 
for these ``particles".

This is in perfect correspondence with what was obtained before for the homogeneous and isotropic cosmology, 
where we worked directly with the scalar field equation of motion. Notice that the analysis of this section
holds if $V (\phi)$ is a  trivial, constant potential or a non-trivial function. 

It is particularly instructive to consider \ref{geodesic5} in the case of an homogeneous and isotropic 
cosmology, in the framework of Friedmann-Robertson-Walker (FRW)
cosmology with the metric taken to be of the form
 \ref{eq11}. In this case, taking also $g_{00}=-1$ and specializing to the case where the scalar field 
depends only on time. In this case the four velocity \ref{geodesic1} gives $u^{\nu}= (1,0,0,0)$ for 
any homogeneous scalar field $\phi$ with non zero time derivative. But if we take the reasonable 
prescription that $u^{\nu}= (1,0,0,0)$ also by continuity, for those isolated points where the scalar 
field $\phi$ has zero time derivative, then all ambiguities get resolved. Indeed, in this homogeneous 
and isotropic cosmological situation,  equation (\ref{geodesic5}) is solved by 
\begin{equation}\label{geodesic7}
\rho _d = constant/a^3
\end{equation}
Recalling the definition for $\rho _d$, this means indeed that (\ref{geodesic7}) represents the 
same solution for $\Phi$ as that found in the previous section (appropriately identifying the 
$constant$ in (\ref{geodesic7}) with constants defined before) and we also see that when 
specialized to the homogeneous and isotropic case produces indeed the required  form for $C(r, \theta)$. 
So we see that the scalar field equation as studied in the previous section is associated with a 
conservation law.

\section{Discussion and conclusion}

In this work we have studied a simple model with a metric measure, $\sqrt{-g}$, and
a non-metric measure, $\Phi$. The model also contained a scalar field with a non-standard,
DBI-like Lagrangian density as given by the third term in \eqref{eq05}. This scalar field Lagrangian 
gave a tachyonic scalar field when the potential satisfied certain conditions. The dynamic
equation for the scalar field given in \eqref{eq09} has an effective potential $\propto [V (\phi ) ]^{-2}$
thus some potentials which would naively lead to tachyonic behavior can result in stable, non-tachyonic
behavior for the two measure model. For example, the potential $[\cosh (\phi)]^{-1}$ which was considered
in \cite{Kutasov2003}, led to a tachyonic scalar field. For the two measure model discussed above this
potential would become $[\cosh (\phi)]^2$ and lead to a regular, stable scalar field.  
 
There are two other noteworthy features of the two measure model detailed above. First, for fairly generic
potentials, $V(\phi )$, the total energy density \eqref{eq23} has a constant term, $M$, plus an ordinary matter 
term, $\frac{D}{M a^3}$. The pressure \eqref{eq24} is negative $-M$. This leads to an equation of state
$w=\frac{p}{\rho}$ which is negative at early times (i.e. when $a(t) \rightarrow 0$) 
and approaches $w=-1$ at late times (i.e. for $a(t) \rightarrow \infty$). 
Second, the scale of the energy density and pressure is determined not by $V(\phi )$, but $M$. 
$M$ is a constant, but from its definition $M=V(\phi ) \sqrt{1- \dot{\phi}^{2}}$ we see that $0<M< V(\phi )$, 
so that the vacuum energy density, $M$, is always lower than $V(\phi )$, and can be as small as we want. 
Furthermore, the limit $M\rightarrow 0$ has clear physical meaning, it is achieved when the scalar 
field $\phi$ approaches its maximum speed ($\dot{\phi}^{2}= 1$).

The two measure model still does not address the ``cosmological coincidence problem" -- i.e.
why the dark energy density and ordinary matter density (including dark matter) are of the
same order of magnitude at the present time. One way to have the two measure model
address this issue is to make the constant of integration $M$ dynamical.
This can be done by considering a 2-brane creation as follows: define $\Phi$ in terms of
a rank-three tensor as $\Phi=\varepsilon^{\mu\nu\alpha\beta}\partial_{\mu}A_{\nu\alpha\beta}$.
The equation of motion with respect to $A_{\nu\alpha\beta}$ still gives that $\frac{K}{\sqrt{-g}}$ is a constant.  
We now couple this $A_{\nu\alpha\beta}$ to a $2$-brane 
\begin{equation}\label{eq25}
\lambda\int A_{\nu\alpha\beta}\mathrm{d}x^{\nu}\wedge\mathrm{d}x^{\alpha}\wedge\mathrm{d}x^{\beta} ~.
\end{equation}  
We assume that on one side of the brane we have $\frac{K}{\sqrt{-g}}=C_{1}$ and on the other side $\frac{K}{\sqrt{-g}}=C_{2}$, where 
$C_{2}-C_{1}=\lambda$. As a result, $M$ could be made dynamical if membrane creation is taken into account. 
The same thing can be done with a $4$-scalar field with $A_{\nu\alpha\beta}$ being a composite 
of the auxiliary scalars, $\varphi ^i$, defined as
\begin{equation}\label{eq26}
A_{\nu\alpha\beta}=\varepsilon_{ijkl}\varphi^{i}
\partial_{\nu}\varphi^{j}\partial_{\alpha}\varphi^{k}\partial_{\beta}\varphi^{l}.
\end{equation} 
A 2-brane creation process to address the cosmological constant problem was suggested in \cite{Brown1987}. In this
work, at late times, a small vacuum energy density was achieved, but with no matter. In our model above
we notice that by lowering the effective vacuum energy density, which is proportional to $M$, we also raise the 
amount of effective dark matter,  which is proportional to $\frac{1}{M}$. Thus lowering of the cosmological constant 
should not result in an empty Universe in our case. Instead the process of brane creation sketched above
would drive the universe to a balance between dark energy and dark matter.      

Another possible approach is that developed in the follow-up paper \cite{Ansoldi-Guendelman}, 
where an additional scalar field was introduced that could produce energy-momentum exchange 
between the vacuum and dust sector.

One should also look at the theoretical justifications for the $\Phi$-term in \eqref{eq05}. 
In this respect it is interesting to note that $\Phi$ appears naturally from the Polyakov action of a 
$3$-brane, with $\varphi^{a}$ being its coordinates, moving in a $4$-D space \cite{Guendelman_Kaganovich2006,Floreanini1990}.

\begin{acknowledgments}
This work has been supported by the Thailand Center of Excellence in Physics.  We thank David Eichler,
Salvatore Capozziello, Gia Dvali and Subash Rajpoot for useful conversations and Neven Bilic for pointing out some important references. 
We would like to acknowledge with thanks for Department of Physics, California State University, Fresno, USA for being host. 
The work of DS has been supported via a 2012 Deutscher Akademischer Austausch Dienst(DAAD) grant.
\end{acknowledgments}

\end{document}